# Meaning of Interior Tomography

*Ge Wang [1] and Hengyong Yu[2,3]*

1. *Biomedical Imaging Cluster, Department of Biomedical Engineering, Rensselaer Polytechnic Institute, Troy, NY 12180, USA;*
2. *Department of Radiology, Wake Forest University Health Sciences, Winston-Salem, NC, 27157, USA;*
3. *Biomedical Imaging Division, VT-WFU School of Biomedical Engineering and Sciences, Wake Forest University Health Sciences, Winston-Salem, NC, 27157, USA*

ge-wang@ieee.org; hengyong-yu@ieee.org

**Abstract:** The classic imaging geometry for computed tomography is for collection of un-truncated projections and reconstruction of a global image, with the Fourier transform as the theoretical foundation that is intrinsically non-local. Recently, interior tomography research has led to theoretically exact relationships between localities in the projection and image spaces and practically promising reconstruction algorithms. Initially, interior tomography was developed for x-ray computed tomography. Then, it has been elevated as a general imaging principle. Finally, a novel framework known as "omni-tomography" is being developed for grand fusion of multiple imaging modalities, allowing tomographic synchrony of diversified features.

**Keywords:** Biomedical imaging, computed tomography, local tomography, interior tomography, omni-tomography

## I. Introduction

The revolutionary work on x-ray computed tomography (CT) by Hounsfield [1] and Cormack [2] has produced a profound impact on the field of medical imaging. Over the past decade, the development of spiral fan-beam/multi-slice/cone-beam CT technologies has dramatically increased and continues increasing the number of CT scans [3, 4].

Smith-Bindman and her colleagues estimated that the number of CT scans was nearly tripled in the USA over the past 15 years, from 52 scans per 1,000 patients in 1996 to 149 scans per 1,000 patients in 2010 [5]. Now, there are about 100 million CT scans annually performed around the world (http://www.buzzle.com/articles/cat-scan-cost.html).

The mathematical twist of spiral cone-beam CT comes from longitudinal data truncation. Because of the beam divergence, spiral cone-beam CT cannot be simplified into fan-beam reconstruction through longitudinal data interpolation like what was done for spiral fan-beam CT [6]. The first spiral cone-beam CT algorithm was derived by Wang et al. [7] by extending the popular circular cone-beam CT algorithm by Feldkamp *et al*. [8]. It has taken CT reconstruction researchers about a decade to find theoretically exact solutions [9-13]. More details can be found in [3, 14].

Encouraged by the impressive results from overcoming the longitudinal data truncation, we were very curious about what would happen with the transverse data truncation. This is actually the well-known "interior problem" [15], in which an internal region of interest (ROI) is irradiated with x-rays only through the ROI to recover the ROI image exactly. Unfortunately, it was already proven long time ago that the interior problem had no unique solution in general [15]. The historical fact that reliable image reconstruction cannot be performed from truncated projection data has contributed to the current CT architectures whereby the detectors are sufficiently wide to cover a transaxial section fully. When a traditional CT algorithm is applied for local reconstruction from truncated data, quantitative accuracy is lost in a reconstructed image, compromising its diagnostic value significantly.

To overcome the transverse data truncation innovatively, with our collaborators we have been developing interior tomography since 2007 [16-26]. Independent results in this area were also performed by peers [27-31]. By interior tomography we mean the theoretically exact yet numerically stable solution to the long-standing interior problem assuming appropriate yet practical *a priori* knowledge (Figure 1). In this sense, it is different from either traditional local tomography that settles with an approximate solution over an ROI [32] or well-known lambda tomography that only captures significant changes in an ROI [33]. Various kinds of *a priori* knowledge are possible to enable interior tomography, such as a known sub-region inside an ROI [16, 27], outside an ROI [31, 34], or a sparsity model of an ROI [21, 24].

A misconception is occasionally heard that approximate local reconstruction is only one constant away from the truth. According to page 169 of Natterer's classic textbook [15], the ambiguity is an infinitely differentiable function whose variation is relatively small well within an ROI but it increases rapidly towards the boundary of the ROI and is not bounded over the ROI. The magnitude estimates in Table 4.1 on page 170 of the same book [15] are ~2-13% for a sub-region of an ROI, which are considered rather significant because the CT number variation of soft tissues is in this range. Actually, we have proven that a non-zero constant ambiguity in an ROI is simply impossible[22], suggesting the non-trivialness of interior tomography.

As an emerging area, interior tomography brings research opportunities and promises practical applications. In this article, we will first review the literature on conventional local tomography that is theoretically approximate, and then discuss the essential meaning of interior tomography in both its special and general forms. The remaining parts of this paper are organized as follows. In section II, we will recall conventional local tomography techniques that can only produce approximate reconstruction from truncated projections. In Sections III and IV, we will respectively describe interior tomography for x-ray CT and other imaging modalities such as SPECT, MRI, and differential phase-contrast tomography. In Section V, we will focus on omni-tomography enabled by the general theory of interior tomography. In Section VI, we will discuss relevant issues and conclude the paper.

## II. Prior Art

The interior problem and approximate solutions were extensively studied in the past decades. These results can be divided into two classes: (1) approximate algorithms for image reconstruction over an ROI [32, 35-42] and (2) lambda tomography algorithms for edge identification within an ROI [33, 43-56], either of which only assumes that local projection data are available. To put interior tomography in perspective, in this section we briefly touch upon these classes.

### II.1. Approximate Local Reconstruction

Since many applications involved the interior problem but it was believed that there would be no unique solution in such cases [15, 38], a number of researchers developed various approximate local reconstruction algorithms for this purpose. For example, Louis and Rieder derived the consistency conditions for the divergent beam transform and studied the singular value decomposition, which showed that the high angular frequency components can be well determined [40]. Sahiner and Yagle combined the non-uniform sampling and circular harmonic expansions for ROI reconstruction [41]. Using wavelets to localize the Radon transform, Rashid-Farrokhi *et al.* developed a multi-resolution method for ROI reconstruction from almost completely local projections [42]. Wiegert *et al.* extrapolated truncated ROI projections from forwarded projections of a previously acquired reference image and obtained excellent ROI images [39].

Our group published multiple papers in this regard. We studied the wavelet local tomography [57-59] for parallel and fan-beam geometries, and applied the results to cone-beam data [60]. Our wavelet filtering and local reconstruction techniques were applied to metal artifact reduction (MAR) [61, 62]. In a local ROI reconstruction scheme we proposed [63], a normal radiation dose is delivered to an ROI while a low radiation dose is applied outside the ROI. If both low and high quality datasets are acquired, we can combine them for excellent local image reconstruction such as for clinical micro-CT possibilities [64]. It is underlined that these local reconstruction algorithms we developed, along with similar algorithms other investigators developed [65-70], were approximate in principle. Hence, they have no theoretical basis to produce exact reconstruction results.

## II.2. Lambda Tomography

In contrast to approximate local reconstruction algorithms, lambda tomography techniques target boundaries only. Let **x** and $\xi$ represent two-dimensional (2D) vectors, $f(\mathbf{x})$ a 2D bounded function, $\hat{f}(\xi)$ the corresponding Fourier transform, we have

$$\begin{cases} \hat{f}(\xi) = \int_{\mathbb{R}^2} f(\mathbf{x}) e^{-i\mathbf{x}\cdot\xi} d\mathbf{x} \\ f(\mathbf{x}) = \dfrac{1}{(2\pi)^2} \int_{\mathbb{R}^2} \hat{f}(\xi) e^{i\mathbf{x}\cdot\xi} d\xi \end{cases}, \qquad (2.2.1)$$

where $\mathbf{R}^2$ denotes the 2D space. Let $\Lambda$ be the Calderon operator defined as

$$\widehat{\Lambda f}(\xi) = \|\xi\| \hat{f}(\xi). \tag{2.2.2}$$

Lambda tomography is to reconstruct the gradient-like function $\Lambda f(\mathbf{x})$ only from directly involved truncated projection data, which was first proposed in [43].

Despite its non-quantitative nature, lambda tomography has a major mathematical advantage over approximate local tomography. Thanks to the simplified reconstruction requirement, lambda tomography eliminates non-local filtering, and image reconstruction can be done with high computational efficiency [33]. Faridani *et al.* studied the properties of the Calderon operator and its adjoint, whose linear combination was used for local reconstruction [45, 46]. Katsevich proposed an efficient numerical scheme to compute lambda tomography from a generalized Radon transform [47], and extended it for limited-angle [48], non-smooth attenuation [50], cone-beam [49] tomography. Furthermore, Katsevich proposed an improved version of cone-beam lambda tomography [54], updating his previous results [49].

Inspired by the results on exact CT reconstruction from data acquired along any smooth scanning curve [3, 71], a theoretically exact fan-beam lambda tomography formula was proved for data collected along a smooth trajectory in terms of the 2D Calderon operator [72]. It was also extended to cone-beam lambda tomography [73] and even with discontinuous trajectories [74]. Compared with the result in [54], the major result in [73] is a special case of [54] with a specific weighting function derived from the 2D exact formula in [72]. Generally, it is impossible to reconstruct a 3D lambda tomography image exactly with a spiral cone-beam scan in terms of the 3D Calderon operator [74]. Hence, a skew cone-beam lambda tomography method was proposed [75]. Recently, Quinto *et al.* reported on electron lambda tomography [51, 76-78]. Katsevich studied motion compensated local tomography [52, 55, 56].

## III. Special Interior Tomography

The development of CT theory is to use an increasingly smaller amount of data for theoretically exact image reconstruction. Examples include half-scan fan-beam reconstruction [79], 2D two-step Hilbert method [80, 81], and Tam-window-based cone-beam reconstruction [9, 82]. Each relaxation of the data condition is a theoretical advancement. In the classic CT literature, a minimum dataset consists of transversely non-truncated projections. Such a global dataset permits theoretically exact reconstruction of a cross-section or an entire volume. In contrast, interior tomography offers an exact inner vision over an ROI only from directly related local projection data, demonstrating in an essential sense and for the first time the uniqueness and stability of local reconstruction of reconstruction of an ROI only from data that directly involve the ROI (Figure 2). This decomposition has sharpened our understanding of the relationship between the image and projection domains, which shows the locality correspondence across these domains.

This aforementioned relationship is fundamentally related to Gel'fand-Graev theory [14, 83]: *Backprojection of differentiated projection data over a PI-line segment gives the truncated Hilbert transform over it*. Here, the PI-line segment is defined as a line-segment connecting two different points on a scanning trajectory, and the backprojection is performed over the data acquired along the corresponding PI-arc (the portion of the scanning trajectory between the two points of the PI-line segment). Let $H_0$ be a set of oriented 1D lines in the 3D real space $\mathbb{R}^3$ through the origin. Since an oriented line has a direction, $H_0$ can be viewed as the unit sphere $S^2$ in $\mathbb{R}^3$ and $\boldsymbol{\alpha} \in S^2$ defines the oriented line $t\boldsymbol{\alpha} \in H_0$, $t \in \mathbb{R}$. Let us fix a 1D oriented curve $\boldsymbol{\gamma}$, which can be represented as an oriented curve $C_\gamma \subset S^2$. For a compactly supported smooth function $f(\mathbf{x})$, let us define an operator $J_\gamma$ on the integral transform

$$\varphi(\boldsymbol{\alpha}, \boldsymbol{\beta}) = Rf(\boldsymbol{\alpha}, \boldsymbol{\beta}) = \int_{\mathbb{R}} f(t\boldsymbol{\alpha} + \boldsymbol{\beta}) dt, \boldsymbol{\alpha} \in S^2, \boldsymbol{\beta} \in \mathbb{R}^3, \qquad (3.0.1)$$

by

$$\left(J_\gamma \varphi\right)(\mathbf{x}) = \frac{1}{2\pi i} \int_{C_\gamma} \sum_{j=1}^{3} \left.\frac{\partial \varphi(\boldsymbol{\alpha}, \boldsymbol{\beta})}{\partial \beta_j}\right|_{\boldsymbol{\beta}=\mathbf{x}} d\alpha_j, \qquad (3.0.2)$$

where $\alpha_j$ and $\beta_j$ represent the $j^{th}$ component of $\boldsymbol{\alpha}$ and $\boldsymbol{\beta}$, respectively. When $C_\gamma$ is a smooth curve on $S^2$ whose end points are diametrically opposite on $S^2$, it is called a quasicycle. Gel'fand and Grave's major results can be re-stated as follows.

*Theorem 3.0.1.* [14, 83] *If* $\gamma$ *is a quasicycle in* $H_0$, *then* $\left(J_\gamma R\right)^2 = c(\gamma)^2 E$ *with* $E$ *being the identity operator and* $c(\gamma)^2$ *a constant. Thus, for the integral transform* $\varphi = Rf$, *one has the inversion formula*

$$J_\gamma R J_\gamma \varphi = c(\gamma)^2 f . \qquad (3.0.3)$$

In the CT field, $J_\gamma R$ is exactly the Hilbert transform relationship between the backprojection data and the original image on a PI-line, which was re-discovered 13 years later [81]; see [14, 83] for more details. In contrast, special interior tomography means that a truncated Hilbert transform data can be uniquely and stably inverted under various quite general assumptions on an ROI, which is a step forward relative to inversion of the finite Hilbert transform and will be explained in the following two sub-sections.

### III.1 Known Sub-region Based Interior Tomography

The first working assumption for interior tomography is the availability of a known sub-region in an ROI [16, 27, 84]. In many scenarios, a sub-region is indeed known in advance, such as air in airways, blood through vessels, or images from prior scans. Our work [16] was inspired by Defrise *et al.*'s work on the partial PI line based Hilbert transform inversion [85]. Initially, we struggled to perform exact local reconstruction assuming one known point in an ROI (Figure 3). This attempt was to solve a conjecture posed in our R01 proposal (EB002667) that Katsevich' helical cone-beam reconstruction [9] can be done along a bundle of curves within the Tam window. We were not successful with interior tomography until a known point assumption was enhanced with a known sub-region of non-zero measure. By the analytic continuation, it was demonstrated that the interior problem can be exactly and stably solved if a sub-region in the ROI is known [16-18]. The key issue for exact interior reconstruction based on a known sub-region is the inversion of a truncated Hilbert transform. A projection onto convex sets (POCS) method and a singular value decomposition (SVD) method were respectively adapted to reconstruct interior images from truly truncated data, and produced excellent results [16, 23, 86]. Although the POCS method is computationally expensive, it is easier to employ additional constraints. The SVD method is computationally much more efficient than the POCS method.

The major theoretical results can be summarized as follows:

**Theorem 3.1.1** [16]: Let $c_1, c_2, c_3, c_4, c_5$ be five real numbers with $c_1 < c_3 < c_5 < c_4 < c_2$. A smooth function $f(x)$ supported on $[c_1, c_2]$ can be exactly reconstructed on $[c_5, c_4)$ if (i) $f(x)$ is known on $(c_3, c_5)$; (ii) $g(x)$ is known on $(c_3, c_4)$, which is truncated data of the Hilbert transform of $f(x)$, and (iii) the constant $C_f$ is known, which is the integral of $f(x)$.

The statement of Theorem 3.1.1 is for a 1D PI-line in a 2D or 3D object through a known sub-region, where $x$ represents the 1D coordinate along the PI-line, $[c_1, c_2]$ represents the intersection between the compact object support and the PI-line, $[c_3, c_4]$ denotes the intersection between the ROI/VOI and the PI-line, $[c_3, c_5]$ denotes the intersection between the known sub-region and the PI-line, $g(x)$ can be calculated from projections collected along the corresponding PI-arc, and $C_f$ can be obtained from the x-ray path containing the PI-line. With the analytic continuation of both $g(x)$ and $f(x)$, a more general result can be expressed as:

**Theorem 3.1.2** [27, 31]: Let $c_1, c_2, c_3, c_4, c_5, c_6$ be six real numbers with $c_1 < c_3 < c_5 < c_6 < c_4 < c_2$. A smooth function $f(x)$ supported on $[c_1, c_2]$ can be exactly reconstructed on $[c_6, c_4)$ if (i) $f(x)$ is known on $(c_3, c_6)$; (ii) $g(x)$ is known on $(c_3, c_5)$ and $(c_6, c_4)$, and (iii) the constant $C_f$ is known.

## III.2. Sparsity-model-based Interior Tomography

The known-sub-region-based interior tomography approach has serious limitations. There are situations where no precise information is available on any sub-region, especially in the scenario of contrast-enhanced CT studies. To address this challenge, the second working assumption we figured out is that an ROI is piecewise constant or piecewise polynomial [21, 22, 24-26, 87]. Then, the total variation (TV) minimization or high order TV (HOT) minimization can be coupled with the data discrepancy minimization to solve the interior problem uniquely and stably [26]. The heuristics behind the TV or HOT minimization is that any "*ghost*" function invisible from measured local data would induce higher TV or HOT values. Hence, such a ghost cannot survive the TV or HOT minimization. Further extensions are possible, such as with a more effective transform based compressibility. Note that the piecewise parametric assumption is already a quite general image model [77], somehow comparable to the limited bandwidth for digital signal processing.

For piecewise constants or piecewise polynomial functions, it is natural to use the space of functions of bounded variation to capture the discontinuities. Mathematically, the total variation of an image $f(x_1, x_2)$ inside an ROI can be evaluated with

$$f_{tv} = \int_{\Omega_a} |f_\nabla(x_1, x_2)| dx, \qquad (3.2.1)$$

where $f_\nabla(x_1, x_2)$ represents a sparsifying transform of $f(x_1, x_2)$ defined on the ROI $\Omega_a$ and $a$ denotes the radius of the ROI. For the commonly used gradient transform in the biomedical imaging field, we can use

$$f_\nabla(x_1, x_2) = \sqrt{\left(\frac{\partial f(x_1, x_2)}{\partial x_1}\right)^2 + \left(\frac{\partial f(x_1, x_2)}{\partial x_2}\right)^2}, \qquad (3.2.2)$$

which is the gradient magnitude or the maximum directional derivation at $(x_1, x_2)$. When there exists an artifact image $f_a(x_1, x_2)$ due to missing data outside an ROI of radius $a$, the total variation becomes

$$\tilde{f}_{tv} = \int_{\Omega_a} \sqrt{\left(\frac{\partial f(x_1, x_2)}{\partial x_1} + \frac{\partial f_a(x_1, x_2)}{\partial x_1}\right)^2 + \left(\frac{\partial f(x_1, x_2)}{\partial x_2} + \frac{\partial f_a(x_1, x_2)}{\partial x_2}\right)^2} dx, \qquad (3.2.3)$$

First, we have the following theorem:

***Theorem 3.2.1** [21, 22, 87]: Suppose that all the projections through an interior ROI of a compactly supported and square integrable nonzero function f are available. The attenuation coefficient inside the ROI can be exactly determined by minimizing the total variation defined by Eq. (3.2.1) from the measured projections if f in the ROI can be decomposed into finitely many constant sub-regions.*

Then, we have the following more general theorem in terms of the high-order total variation (HOT) [24]:

***Theorem 3.2.2**[24]: Suppose that all the projections through an interior ROI of a compactly supported and square integrable nonzero function f are available. The attenuation coefficient inside the ROI can be exactly determined by minimizing the high-order total variation (HOT) from the measured projections if f in the ROI can be decomposed into finitely many sub-regions where the function can be modeled as a polynomial, with the HOT being defined as*

$$\text{HOT}_{n+1}(f) = \sum_{i=1}^{m} \sum_{j>i, j \in N_i} \int_{\Gamma_{i,j}} |f_i - f_j| ds + \int_{\Omega_a} \min\left\{\sqrt{\sum_{r=0}^{n+1}\left(\frac{\partial^{n+1} f}{\partial x_1^r \partial x_2^{n+1-r}}\right)^2}, \sqrt{\left(\frac{\partial f}{\partial x_1}\right)^2 + \left(\frac{\partial f}{\partial x_2}\right)^2}\right\} dx. \quad (3.2.4)$$

In Eq. (3.2.4), $f(x_1, x_2)$ is a piecewise $n^{th}$ ($n \geq 1$) order polynomial function in the ROI, $m$ is the number of sub-regions, $\Gamma_{i,j}$ represents piecewise smooth boundaries between $i^{th}$ and $j^{th}$ sub-regions, and $N_i$ is the neighborhood regions of the $i^{th}$ sub-region.

### III.3. Practical Performance

To demonstrate the application of interior tomography, a cardiac CT study was performed on a state-of-the-art GE discovery CT750 HD scanner at Wake Forest University Health Sciences [88]. After appropriate pre-processing, we obtained a 64-slice fan-beam sinogram. The radius of the scanning trajectory was 53.852 cm. Over a 360° range, 2200 projections were evenly acquired. For each projection, 888 detector elements were equi-angularly distributed, which defined a field of view of 24.92 cm in radius. From each sinogram, an 800x800 image matrix was reconstructed using the filtered backprojection (FBP) method as a benchmark.

To evaluate our sparsity-model-based interior tomography techniques, each projection was truncated by discarding 300 detector elements on each of its two sides to just cover an interior ROI of 8.70 cm in radius. This interior reconstruction process was implemented in a soft-threshold filtering framework, with a flowchart in Figure 4 [89]. The algorithm included two major steps. In the first step, the ordered-subset simultaneous algebraic reconstruction technique (OS-SART) [90] was used to reconstruct a digital image from all the truncated projections. The initial image was set to zero, and the 2200 truncated projections were divided into 88 subsets, and each of them includes 25 uniformly distributed projections. Let the subset index be as 1, 2,...., 88 sequentially. In each iteration, all the subsets were employed in the order of 1, 45, 2, 46, ..., 44, 88 to suppress artifacts. Then, the total difference of an intermediate image was minimized using the fast iterative soft thresholding algorithm (FISTA) [91] and a pseudo-inverse discrete difference transform (DDT) [89]. To accelerate the converging speed, the projected gradient method [92] was used to determine an optimal threshold for the soft thresholding filtration. The two steps were alternated until the refinement in the projection domain became insignificant. The reconstructed image was also in an 800x800 matrix covering a region of 44.86x44.86 cm$^2$. It can be seen in Figure 5 that the difference between the interior reconstruction and global reconstruction became smaller and smaller with the increment of the iteration index, and should approach zero in the limiting case when scattering, beam-hardening and noise are negligible according to our theoretical analysis [21, 26, 87].

## III.4. Advantages and Limitations

Interior tomography allows exact reconstruction from less data. Naturally, there are major advantages from the less-is-more effect in this circumstance. At the beginning of this section, we have shown that less means a deeper theoretical understanding of the local reconstruction mechanism. In the following, we suggest that less means lower, larger, and faster, among other benefits.

**Less is lower** – Less data is equivalent to lower radiation dose, because of not only a narrower beam but also a less angular sampling requirement in the longitudinal studies or multi-scale scenarios. The smaller an ROI is, the narrower the beam will be, and the less radiation gets involved. Because of the narrower beam, the number of scattered photons becomes less, which improves contrast resolution. Hence, one can reduce radiation dose further, given the same contrast resolution as that of the global reconstruction from a complete dataset. Interestingly, for a smaller ROI in a case of a known background outside the ROI, the spatial resolution of global reconstruction can be maintained for interior reconstruction at a less angular sampling rate. Our numerical results are consistent with the heuristics that only two projections (two independent readings per projection, which can be assured by choice of rays) are needed to reconstruct a 2 by 2 ROI, three projections (three independent readings per projection) are sufficient to reconstruct a 3 by 3 ROI, and so on. In other words, the number of projections could be reduced with interior tomography in favorable cases (say, the region outside the ROI is roughly known), accordingly reduced is the radiation dose in this scenario.

Recently, compressive sensing has been a hot topic in the signal processing community. Inspired by this powerful theory, "*few-view*" interior tomography was studied with encouraging results [30, 93-96] (Figure 6). Another dimension towards dose reduction is to perform statistical interior tomography [97, 98], which is an adapted version of conventional statistical reconstruction [99]. There is no doubt that less data and less radiation mean lower cost and safer CT systems. Efforts in this direction will be significant for either "*low-end*" or "*high-end*" CT systems.

**Less is Larger –** Acquisition of less projection data is achieved with a narrower beam, and an object larger than the beam width is not a concern anymore. In other words, interior tomography can handle objects larger than a field of view. This flexibility can certainly enhance the utility of a CT scanner. In nano-CT studies, one reduces a sample into a narrow x-ray beam for complete projection profiles. In this tedious process, morphological and functional damages may be induced. Supported by an NSF/MRI grant and in collaboration with Xradia, we are developing a next generation nano-CT system capable of focusing on an ROI and reconstructing it accurately within a large object (Figure 7). In a geo-science project, fossils from the Ediacaran Doushantuo Formation (ca. 551-635 million years old) were investigated, which are too valuable to be broken, and demand interior tomography. Another example is the large patient problem, i.e., a patient is larger than the field of view of a CT scanner, which can now be solved using theoretically exact interior tomography, instead of using conventional approximate methods such as extrapolation for data completion.

**Less is faster –** Interior tomography allows a smaller detector size, a faster frame rate, and more imaging chains in a gantry space, all of which contribute to an accelerated data acquisition process. Clearly, the evolution of the CT technology is for faster and faster scanning. Spiral CT is revolutionary [3] because it enables truly volumetric and dynamic tomographic x-ray imaging, utilizing a high-tech slip ring, a modern detector array, sophisticated reconstruction methods and system architectures. The scanning speed of more than 3 turns per second has reached the mechanical upper bound. While dual-source scanners are well received, the scanning speed is still not sufficient to handle challenging cases of rapid or irregular heart beating and other physiological processes. The physical obstacle to use more sources and detectors is the rather limited gantry space where at most we can assemble a triple-source system [100]. An interior tomographic imaging chain is slimmer, weighs lighter, facilitates faster rotation, records less data, and transfers them more rapidly. Thus, multiple local imaging chains can be fitted into a gantry for simultaneous data acquisition and ultrafast interior reconstruction [20].

In 2009, our group proposed a multi-source interior tomography scheme to improve temporal resolution of CT [20] (Figure 8). For a CT scan of the heart, a field of view of ~15cm in diameter is preferred (apart from some very special cases where only part of the heart or coronary artery would be of interest) and can be achieved using multi-source interior tomography by scanning the source-detector chains in a circular, saddle-curve or another trajectory. In the case of $2N+1$ sources, the radius of the field of view can be maximized by arranging the source-detector pairs equi-angularly, or re-configured for other purposes. This scheme promises to revitalize the dynamic spatial reconstructor (DSR) concept [101, 102], and in the limiting case allows instantaneous temporal resolution. However, the cross-scattering effects between image chains are challenging for the multi-source scheme, and must be managed such as with algorithmic compensation or beam multiplexing.

In the CT field, "CT" and "*scanning*" have been together and inseparable. Nevertheless, with the many-source interior tomography scheme, taking a snapshot of an ROI is feasible without any scanning. When such a snapshot does not adequately cover features of interest, one can always take another snapshot. In other words, one can roam within an object without scanning. Although these snapshots are not taken at the same time, each of them is an instant reflection of the ROI at the corresponding moment, and could give critical information for medical and non-medical studies.

**Less is not always better** – While compressive sensing inspired reconstruction algorithms produce visually pleasing images, diagnostically critical information may be hidden or lost [103]. This potential risk is substantial with interior tomography especially few views. In practice, the minimum amount of data can only be determined in a task specific fashion, through extensive tests, and with an optimized algorithm. Whenever the information carried by data is less than the number of degrees of freedom of the sparsest model for an ROI, the solution will not be unique, and the promise of interior tomography cannot be hold. This pre-caution must be underlined to avoid any adverse effect. Even if sufficiently many views are available for interior tomography, the stability of interior reconstruction is not as good as that of global reconstruction for a well-known mathematical reason [26]. This weakness can be addressed with more prior knowledge, sparsely sampled more global data, or other means.

## IV. General Interior Tomography

Interior tomography has been extended from the CT field to other tomographic imaging modalities, such as single-photon emission computed tomography (SPECT) [104-110], MRI [111, 112], differential phase-contrast tomography [113], and spectral CT [98]. Based on these established interior tomographic imaging modes, we have postulated the following abstraction that *theoretically exact local tomography can be always done only from indirect measured data that directly involves a region of interest under rather general assumptions*. This general interior tomography principle can be further appreciated in the following case studies.

### IV.1. Interior SPECT

While the CT reconstruction algorithms are being advanced rapidly, SPECT techniques are being improved as well [114-118]. As a unique tomographic imaging modality, SPECT is to reconstruct a radioactive source distribution within a patient from data collected with a gamma camera at multiple orientations. Different from the line integral model for x-ray CT, a SPECT projection can be mathematically modeled as an exponentially attenuated Radon transform [118, 119]. Thus, the CT reconstruction is a special case of SPECT when the attenuation coefficients are set to zero, but CT reconstruction techniques cannot be used for SPECT in general.

Encouraged by excellent results from special interior tomography that was formulated for CT, interior SPECT was first developed assuming a constant attenuation background [104]. Based on interior tomography results in the CT field, we proved in 2008 that theoretically exact interior SPECT is feasible from uniformly attenuated local projection data, aided by a known sub-region in an ROI [104]. With total variation minimization techniques [21, 24], we further proved that if an ROI is piecewise polynomial, then it can be uniquely reconstructed from truncated SPECT data that go directly through the ROI [105]. Based on the above results, other researchers developed variants of interior SPECT for more flexibility [106-109].

Let $f(\mathbf{x})$ be a 2D smooth distribution function on a compact support $\Omega$ with $\mathbf{x} = (x, y) \in \Omega$. In a parallel-beam geometry, a SPECT projection of $f(\mathbf{x})$ is [118]:

$$P_o(\theta, s) = \int_{-\infty}^{\infty} f(s\boldsymbol{\theta} + t\boldsymbol{\theta}^{\perp}) e^{-\int_{t}^{\infty} \mu(s\boldsymbol{\theta} + \tilde{t}\boldsymbol{\theta}^{\perp}) d\tilde{t}} dt, \quad (4.1.1)$$

where the subscript "$o$" denotes original projection data, $\boldsymbol{\theta} = (\cos\theta, \sin(\theta))$, $\boldsymbol{\theta}^{\perp} = (-\sin\theta, \cos(\theta))$, and $\mu(\mathbf{x})$ the attenuation map on the whole support. A practical uniform attenuation map can be defined as

$$\mu(\mathbf{x}) = \begin{cases} \mu_0 & \mathbf{x} \in \Omega \\ 0 & \mathbf{x} \notin \Omega \end{cases}, \quad (4.1.2)$$

where $\mu_0$ is a constant. Since the object function is compactly supported, we can determine the length of the intersection between the support $\Omega$ and the integral line for $P_o(\theta, s)$. Without loss of generality, we denote this length as $t_{\max}(\theta, s)$, and Eq. (4.1.1) becomes

$$P_o(\theta, s) = e^{-\mu_0 t_{\max}(\theta, s)} \int_{-\infty}^{\infty} f(s\boldsymbol{\theta} + t\boldsymbol{\theta}^{\perp}) e^{\mu_0 t} dt. \quad (4.1.3)$$

Let us assume that the compact support $\Omega$ and the constant coefficient $\mu_0$ are known. By multiplying a weighting factor $e^{\mu_0 t_{\max}(\theta, s)}$, the projection model for SPECT is reduced to

$$P_w(\theta, s) = P_o(\theta, s) e^{\mu_0 t_{\max}(\theta, s)} = \int_{-\infty}^{\infty} f(s\boldsymbol{\theta} + t\boldsymbol{\theta}^{\perp}) e^{\mu_0 t} dt, \quad (4.1.4)$$

where the subscript "w" indicates weighted projection data.

Let us denote a weighted backprojection of the differential projection data as

$$g(\mathbf{x}) = \int_0^{\pi} e^{-\mu_0 \mathbf{x} \cdot \boldsymbol{\theta}^{\perp}} \left. \frac{\partial P_w(\theta, s)}{\partial s} \right|_{s = \mathbf{x} \cdot \boldsymbol{\theta}} d\theta. \quad (4.1.5)$$

In 2004, Rullgard proved a relationship linking an object image $f(\mathbf{x})$ and the weighted backprojection $g(\mathbf{x})$ as [118]

$$g(\mathbf{x}) = g(x,y) = -2\pi PV \int_{-\infty}^{+\infty} ch_{\mu_0}(y-\tilde{y})f(x,\tilde{y})d\tilde{y}, \qquad (4.1.6)$$

where "PV" represents the Cauchy principle value integral, and $ch_{\mu_0}$ is defined as

$$ch_{\mu_0}(y) = \frac{\cosh(\mu_0 y)}{\pi y} = \frac{e^{\mu_0 y} + e^{-\mu_0 y}}{2\pi y}. \qquad (4.1.7)$$

When $\mu_0 \to 0$, $ch_{\mu_0}$ becomes the Hilbert transform kernel, in consistence with the results in the CT field [80]. Eq. (4.1.7) is called a generalized Hilbert transform.

We have two important theorems for interior SPECT, corresponding to the known-sub-region-based and sparsity-model-based special interior tomography approaches respectively.

**Theorem 4.1.1** *[104]: Let $c_1, c_2, c_3, c_4, c_5$ be five real numbers with $c_1 < c_3 < c_5 < c_4 < c_2$. A smooth function $f(x)$ supported on $(c_1, c_2)$ can be exactly reconstructed on $(c_5, c_4)$ if (i) $f(x)$ is known on $(c_3, c_5)$, (ii) $g(x)$ is known on $(c_3, c_4)$, which is the truncated data of the generalized Hilbert transform, and (iii) the constant $\mu_0$ and $m_{\mu_0}$ are known, where*

$$m_{\mu_0} = \int_{c_1}^{c_2} f(x)\cosh(\mu_0 x)dx, \qquad (4.1.5)$$

*and $g(x)$ can be computed using Eq. (4.1.5) proved by Rullgard [118].*

In Theorem 4.1.1, all the relevant notations have the same meanings as that in Theorem 3.1.1, and Theorem 4.1.1 can reduce to Theorem 3.1.1 when $\mu_0 = 0$.

**Theorem 4.1.2** *[105]: Suppose that $f_0(\mathbf{x})$ is a smooth function in $\Omega_A$ and $f_0(\mathbf{x})$ is a piecewise n-th ($n \geq 1$) order polynomial function in an ROI $\Omega_a \subset \Omega_A$; that is, $\Omega_a$ can be decomposed into finitely many sub-domains $\{\Omega_i\}_{i=1}^{m}$ such that*

$$f_0(\mathbf{x}) = f_i(\mathbf{x}), \text{ for } \mathbf{x} \in \Omega_i, \ 1 \leq i \leq m, \qquad (4.1.6)$$

*where $f_i(\mathbf{x})$ is an n-th order polynomial function, and each sub-domain $\Omega_i$ is adjacent to its neighboring sub-domains $\Omega_j$ with piecewise smooth boundaries $\Gamma_{i,j}$, $j \in N_i$. All the attenuated projections defined by Eq. (4.1.1) through the ROI are available and the constant attenuation coefficient $\mu_0$ is known. If $h(\mathbf{x})$ is the reconstructed image from all the local attenuated projections, $\text{HOT}_{n+1}(h) = \min_{f=f_0+u_1} \text{HOT}_{n+1}(f)$ where $u_1(\mathbf{x})$ is an analytic function in $\Omega_a$ due to the missing data outside the ROI, and*

$$\text{HOT}_{n+1}(f) = \sum_{i=1}^{m} \sum_{j>i, j\in N_i} \int_{\Gamma_{i,j}} |f_i - f_j| ds + \int_{\Omega_a} \min\left\{ \sqrt{\sum_{r=0}^{n+1}\left(\frac{\partial^{n+1} f}{\partial x_1^r \partial x_2^{n+1-r}}\right)^2}, \sqrt{\left(\frac{\partial f}{\partial x_1}\right)^2 + \left(\frac{\partial f}{\partial x_2}\right)^2} \right\} dx,$$

*(4.1.7)*

*then $h(\mathbf{x}) = f_0(\mathbf{x})$ for $\mathbf{x} \in \Omega_a$ [105].*

To demonstrate the utility of interior SPECT, we downloaded a SPECT cardiac perfusion image from Internet and modified it into a realistic 128 by 128 image phantom, covering an area of 128 by 128 mm$^2$. It represents a radionuclide distribution in a human heart. In our simulation, we assumed a constant attenuating background $\mu_0 = 0.15$ cm$^{-1}$ on a compact support of a standard patient chest size. We used an equi-spatial detector array consisting of 78 detector elements, each of which was 1.0 mm in length. For a full-scan, we acquired 128 equi-angular projections. Clearly, this image phantom does not satisfy the piecewise polynomial model in the case of *n=1*. To improve the stability of interior SPECT, two additional constraints were incorporated into the projection-onto-convex-sets (POCS) framework: (1) non-negativity, which means that the radionuclide distribution should not be negative, and we kept making negative values zero during the iterative process; (2) compactness, which means that the radionuclide distribution should be inside the human body, and we made the pixels outside the body-contour zero. Our results (Figure 9) demonstrate that interior SPECT worked well, even if the piecewise polynomial model was not exactly satisfied.

## IV.2. Interior MRI

A number of important magnetic resonance imaging (MRI) applications require high spatial and temporal resolutions. With technological advances in MRI hardware, multiple echo data acquisition and image reconstruction strategies, fast MRI is a hot topic.

However, the current achievable spatial and temporal resolutions are often insufficient, such as in cases of vulnerable plaque characterization that was attempted for carotid but is not feasible yet for coronary *in vivo*, imaging-guidance biopsies/intravascular procedures, study of brain function, extraction of cancer biomarkers, and etc. It is well known that the existing techniques for improving temporal resolution compromise spatial resolution, with image noise being fixed over a whole field of view. However, if one targets an ROI instead of the whole patient cross-section, both spatial and temporal resolutions can be significantly improved. A key point is that such an interior MRI mode can be implemented with either a global or local background magnetic field. Needless to say, a local background magnetic field will greatly relax the engineering requirement and the system cost.

Given a locally homogeneous main field just enough to cover an ROI, to extract MR signals for this ROI we can use a time-varying gradient method. The idea is to keep a gradient field zero at an ROI slice and change the gradient rapidly off that level. As a result, iso-regions of magnetic strength are incoherently excited outside the ROI, avoiding signals that complicate the ROI imaging. This idea can be extended for volume of interest imaging. Let us take 2D interior MRI as an example. While an ROI is selectively excited, standard linear x and y gradient fields can be applied with slopes $G_x$ and $G_y$ respectively for spatial encoding to produce the ROI signal as follows:

$$s(t) = \int_{ROI} \rho e^{i[(\gamma G_x t)x + (\gamma G_y t)y]} dxdy, \qquad (4.2.1)$$

where $\rho$ represents a 2D MR image to be reconstructed, $\gamma$ is a constant, and $t$ is a 1D parameter for the measured data. Eq. (4.2.1) corresponds to one line through the origin in the k-space. If we rotate the magnetic field over a range of $180°$, all the measured data will fully cover the whole k-space. This continuous model can be discretized and solved using a compressive sensing method [112] (Figure 10). While interior MRI as described here is mathematically the easiest, compressive sensing techniques can still be applied for various well-known benefits.

## IV.3. Interior Differential Phase Contrast Tomography

The propagation of x-rays in a medium is characterized by the complex index of refraction $n = 1 - \delta + i\beta$, where $n$ is approximately 1, $\delta$ and $\beta$ quantify phase shift and attenuation magnitude respectively [120]. The cross section of x-ray phase shift is one thousand times larger than that of linear attenuation in the 20–100keV range. This suggests that phase-contrast imaging has higher sensitivity for light elements than attenuation-contrast imaging [121, 122]. The contrast-to-noise ratio of differential phase contrast CT images is superior to the attenuation-contrast CT counterpart [123]. Therefore, phase-contrast imaging can observe subtle but critical structures of soft biological tissues [124-129]. Moreover, the refractive index of tissues is inversely proportional to the square of the x-ray energy while the absorption coefficient decreases as the fourth power of the x-ray energy [120]. Hence, x-ray phase-contrast imaging is suitable to operate at higher energies (>30keV) for lower radiation dose than attenuation imaging [130, 131]. Higher energy x-ray imaging has an important potential for studies on large animals or patients [132]. Recent work shows that differential phase-contrast CT manifests a noise power spectrum (NPS) with a $1/|k|$ trait, while conventional CT NPS goes with $|k|$. This is a significant difference in noise granularity, favoring differential phase-contrast CT [133-135].

Over the past year, interior differential phase-contrast tomography has been studied. It has been proved that the unique interior reconstruction is guaranteed from truncated *differential* projection data that only go through an ROI. The main theoretical results are as follows [25, 113]:

**Theorem 4.3.1** *[25, 113]: If a refractive index image is known on a small sub-region inside an ROI, then the refractive index function on the ROI can be uniquely and stably determined from the truncated differential phase shift data through the ROI.*

**Theorem 4.3.2** *[25, 113]: If (1) a refractive index image of an object is piecewise polynomial in an ROI, (2) another refractive index image is also piecewise polynomial in the same ROI, and (3) both the images have the same differential phase shift data through the ROI, then the two images are identical.*

**Theorem 4.3.3** *[25, 113]: The high order total-variation (HOT) of a piecewise polynomial refractive index function in an ROI is smaller than that of another function, provided that they have the same differential phase shift through the ROI.*

Theorem 4.3.2 shows that the interior reconstruction of a refractive index function from truncated differential phase shift data has a unique solution within the class of piecewise polynomial functions. Theorem 4.3.3 confirms that if the true refractive index distribution is indeed piecewise polynomial in an ROI, the differential phase-contrast interior reconstruction can be performed through the HOT minimization. Because a common diffractive index distribution can be approximated by a piecewise polynomial function, an ROI can be accurately reconstructed *in principle* using HOT minimization (Figure 11) subject to the data fidelity.

## V. Omni-Tomography

Enabled by the general interior tomography principle, we recently proposed an omni-tomographic imaging strategy, which is also referred to as omni-tomography [112] (Figure 12). The essential points of omni-tomography were highlighted in a talking point article: http://medicalphysicsweb.org/cws/article/opinion/51026. Our team is actively working along this direction, with potential targets including interior-CT-MRI and interior-CT-SPECT combinations.

Multimodality imaging with targeted agents such as multimodality probes holds great potential for early screening, accurate diagnosis, prognostic value, and interventional guidance. PET-CT, SPECT-CT, optical-CT, and PET-MRI are well-known examples of dual-modality successes. In 2009, Cherry with University of California–Davis asked an inspiring question on what would be "*a more general trend towards harnessing the complementary nature of the different modalities on integrated imaging platforms*" [136].

Preclinical and clinical studies critically depend on *in vivo* tomography of diversified features that cannot be provided by a single modality in many cases. Software-based registration of different types of images has been widely reported but it has major limitations, especially when high spatial and temporal resolutions are needed for fast or transient physiological phenomena. Hence, the seamless fusion of all needed imaging modalities would be the Holy Grail. However, such a grand fusion has been challenged by the physical conflicts of these scanners.

To address this grand challenge, omni-tomography lays the foundation for the integration of multiple major tomographic scanners into a single gantry [112]. As such, each of tomographic scanners can be made thinner or smaller, and be fused together for comprehensive and simultaneous data acquisition from an ROI. We believe that this new thinking represents the next stage of multimodality fusion to optimally reveal spatiotemporal links in physiological, pathological and pharmaceutical studies. Additionally, omni-tomography could be cost-effective in terms of equipment space and patient throughput.

Omni-tomography has many potential clinical applications. For example, an interior-CT-MRI system could be fabricated for cardiac and stroke imaging. An interior CT-MRI scanner can target the fast-beating heart for registration of functions and structures, delivery of drugs or stem cells, and guidance of complicated procedures such as heart valve replacement. In [112], we presented a unified interior CT-MRI reconstruction method. This has the potential to reduce radiation dose with MRI-aided interior CT reconstruction. Also, CT-aided interior MRI reconstruction can generate fine details with little motion blurring. It is recognized that the rotating x-ray source and detector can interfere with MRI. A solution is to use a stationary multi-source interior CT scheme [20]. Since interior CT components are now fixed, the electromagnetic shielding for interior MRI becomes much easier. A comprehensive cost-performance analysis will be needed for an interior-CT-MRI prototype, but we expect such a machine will come into reality sooner or later.

## VI. Discussions and Conclusion

As mentioned earlier, interior tomography and compressive sensing can be combined for reconstruction from an even less amount of data, leading to attractive imaging options such as few-view CT. It is well known that few-view CT does not give a unique solution in general, and neither would interior few-view CT. Again, the key is to utilize prior knowledge as much as feasible and achieve a unique solution or at least minimize image artifacts or biases. In this regard, compressive sensing techniques such as dictionary learning could help [93, 137-139] (Figure 13).

It is appropriate to emphasize that less is not always more. While compressive sensing inspired reconstruction algorithms produce visually pleasing images, diagnostically critical information may be hidden or lost [103]. This potential risk is substantial with interior tomography as well. In practice, the minimum amount of data can only be determined in a task specific fashion, through extensive tests, and with an optimized algorithm. Whenever the information carried by data is less than the number of degrees of freedom of the sparsest model for an ROI, the solution will not be unique, and the promise of interior tomography cannot be overly claimed. This pre-caution is necessary to avoid any adverse effect.

We run a multi-scale CT facility, which covers a six order of magnitude range in terms of object size and image resolution (http://www.imaging.sbes.vt.edu/research/sam-ct). An interior imaging capability can be implemented for nano-, micro- and macro-CT to enhance the imaging performance significantly.

In the functional space the classic CT theory assumes [15], a theoretically exact local reconstruction does requires a global dataset in the form of integrals over hyper-planes. However, when the functional space is restricted by an appropriate constraint, such as a known sub-region, piece-wise constant or polynomial functions, a theoretically exact local reconstruction can be indeed performed from a purely local dataset. Generally speaking, types of indirect measurement can be more complicated than integrals over hyper-planes, and associated inverse problems can be similarly posed. Targeting exact and stable interior reconstruction would be a promising direction for research on inverse problems with new types of indirect measurement.

Omni-tomography would serve as the first-of-its-kind platform enabling multi-physics/coupled-physics-based imaging. This type of physical coupling and instantaneous imaging could suggest new imaging modes for synergistic information. Examples are already available in the biomedical imaging field. It has been illustrated that CT and MRI data are synergistic in the image domain [139] (Figure 14). It is underlined here that the intrinsic physical interaction would be more fundamental. For example, photoacoustic imaging combines ultrasound resolution and optical contrast and has been widely used [140]. In this imaging mode, laser pulses are delivered into biological tissues and partially absorbed to generate transient thermo-elastic expansions and ultrasonic waves. This allows a combination of high spatial resolution of ultrasound imaging and high contract resolution of optical imaging, revealing unique physiologic and pathologic information. As another example, temperature-modulated bioluminescence tomography utilizes focused ultrasound to heat a small animal body, modulate bioluminescent light emission in vivo, and improve reconstructed image quality [141]. In the same spirit, temperature-modulated fluorescence tomography utilizes recently emerged temperature sensitive fluorescence contrast agents for better imaging performance [142]. Yet another example uses ultrasound and MRI simultaneously for temperature-change-based thermal tomography [143]. In light of omni-tomography, many inspiring questions can be asked. For example, when CT and MRI scans are performed at the same time, will the imaging contrast mechanisms interfere? The Zeeman affect means splitting a spectral line into several components under a static magnetic field. Hence, it is not unreasonable to guess that x-ray linear attenuation characteristics, especially small-angle scattering coefficients, may be altered by a strong magnetic field. When x-rays go through a patient, three types of scattering happen, including coherent scattering, photoelectric absorption and Compton scattering. It is not impossible that these scattering interactions will make T1 and T2 change slightly which are possibly measurable. It is hoped that the strengths of such modulations depend on tissue types and physiological statuses. If that is the case and measurable, we will be able to see new information that cannot be seen if CT and MRI are sequentially performed. In other words, coupled physics imaging could be the next focus of omni-tomographic research.

Very recently, Zeng and Gullberg pointed out that the non-negativity and piecewise-constant constraints do not guarantee a unique solution to the interior problem if the number of views is finite [144]. This is not surprising since interior tomography theory, like traditional analytic tomography theory, was developed under the assumption of continuous angular sampling, despite the piecewise model introduced over an ROI. When the number of views is finite, the general non-uniqueness of tomographic problems follows [15]. For general tomographic reconstruction without data truncation, the Shannon sampling theorem is applied for practical reconstruction. However, this logic cannot be directly extended to interior tomography, and some additional conditioning should be needed, such as the number of views must be large enough so that the minimal polygon constructed in [144] becomes larger than the whole compact support of an object of interest, and hence can be excluded. The supportive point made in [144] is that even in a case of highly truncated projections down to 5.2% only 30 views are sufficient to remove the polygonal bias [144]. In other words, in almost all the practical applications the polygonal bias does not affect image quality in principle. Clearly, interior tomography theory can be further developed to accommodate this and other constraints while maintaining the theoretical exactness aided by appropriate knowledge [145, 146].

Although interior tomography theory has been developed based on individual imaging modalities (e.g. CT, SPECT, phase-contrast tomography), we believe that the interior tomography principle can be refined for a rigorous unification. Let us consider a general weighted integral over a sub-domain of an object, where the sub-domain is compact with a smooth boundary, and controlled by two parameters. When the first parameter of the sub-domain is fixed, varying the second parameter will move the sub-domain smoothly over an ROI in one direction specified by the first parameter. Specifically, let us assume that the measurement be in the form of a P transform of weighted integrals over sub-domains, where P is a polynomial of differential operators up to a finite order. It is our hypothesis that theoretically exact and stable ROI reconstruction can be done from the generalized measures that directly involve an ROI; that is, (1) the intersection of each sub-domain and the ROI is non-empty, (2) all measures satisfying (1) are available, and (3) the ROI is sparse satisfying the piecewise constant/polynomial model or in another linear transform domain.

It is underlined that interior tomography is a general approach and is extremely attractive in numerous applications, wherever we need to handle large objects, minimize radiation dose, suppress scattering artifacts, enhance temporal resolution, reduce system cost, and increase scanner throughput, as summarized in Table 1. In table 1, Oto CT stands for otolaryngology-oriented CT, NDE for non-destructive examination, OCT for optical computed tomography, and TEM for transmission electron microscopy.

In conclusion, we have reviewed recent tomographic progress in overcoming data truncation, being it longitudinal or transverse and with the latter as the emphasis of this paper. By doing so, we have covered special and general versions of interior tomography. Finally, we have proposed the grand fusion concept leading to omni-tomography. We are excited by the bright future of interior tomography and omni-tomography, and very much interested in theoretical investigation, computational optimization, systematic prototyping, and biomedical applications through interdisciplinary collaboration.

*Table 1. Potential of interior tomography, where "xxx" stands for necessary and/or significant important, "xx" for important, and "x" for useful.*

|  | *Larger Object* | *Less Radiation* | *Lower Scattering* | *Faster Acquisition* | *Smaller Detector* | *Higher Throughput* |
|---|---|---|---|---|---|---|
| *Cardiac CT* | xx | xxx | xxx | x | x | xx |
| *Lung CT* | xx | xxx | xxx | xx | xx | xx |
| *Oto CT* | x | xx | x | x | xxx | xx |
| *Dental CT* | x | xx | x | x | xxx | xx |
| *O-arm CT* | x | xxx | x | x | x | xxx |
| *Micro CT* | xxx | xxx | xx | xxx | xxx | xxx |
| *Nano CT* | xxx | xx | xx | xx | xxx | xx |
| *NDE* | xxx | x | xx | x | xx | xxx |
| *MRI* | x | NA | NA | xxx | NA | xxx |
| *SPECT* | xxx | NA | NA | xxx | xx | xxx |
| *PET* | xxx | NA | NA | xxx | xx | xxx |
| *OCT* | xxx | xx | NA | xxx | xx | xxx |
| *Ultrasound* | xxx | NA | x | xxx | x | xx |
| *TEM* | xxx | x | xx | xxx | xx | xx |

## Acknowledgment

This work was partially supported by the NIH/NIBIB Grant EB011785, the NSF grant MRI/CMMI 0923297, as well as the NSF CAREER Award 1149679. Major collaborators include Drs. Yangbo Ye, Jiangsheng Yang, Ming Jiang, Alexander Katsevich, Wenxiang Cong, Hao Gao, Jie Zhang, Yang Lv, Jun Zhao, Tiange Zhuang, Michael Vannier, Anthony Butler, Philip Butler, Steve Wang, Michael Fesser, Erik Ritman, Deepak Bharkhada, Bruno DeMan, Guohua Cao, Otto Zhou, *et al*.

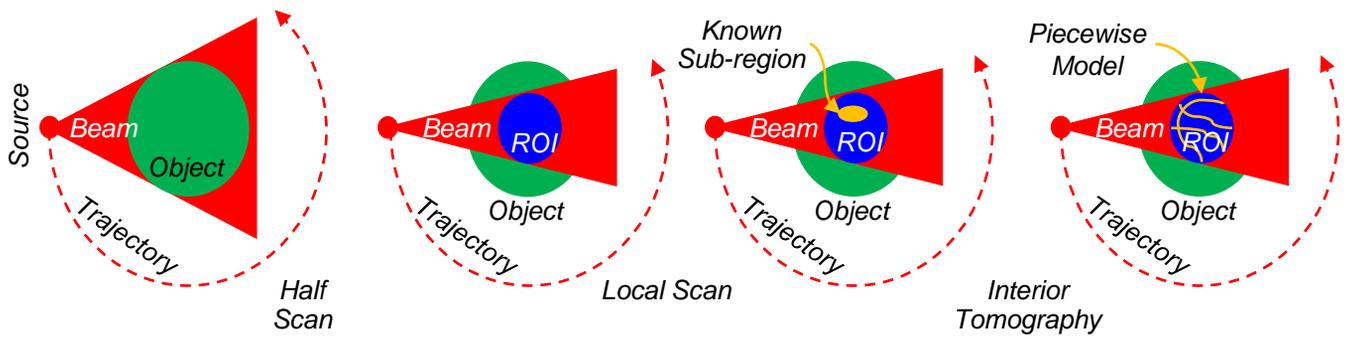

*Figure 1. Interior tomography idea for theoretical exact and stable reconstruction of a region of interest (ROI) only from projective data associated with x-rays through the ROI. (a) Global reconstruction from complete projections (the classic imaging geometry), (b) internal ROI reconstruction from truncated projections (the interior problem) without a unique solution in a constrained setting, (c) and (d) theoretically exact and stable interior reconstruction assuming a known subregion and a sparsity model respectively. This figure is modified from [88].*

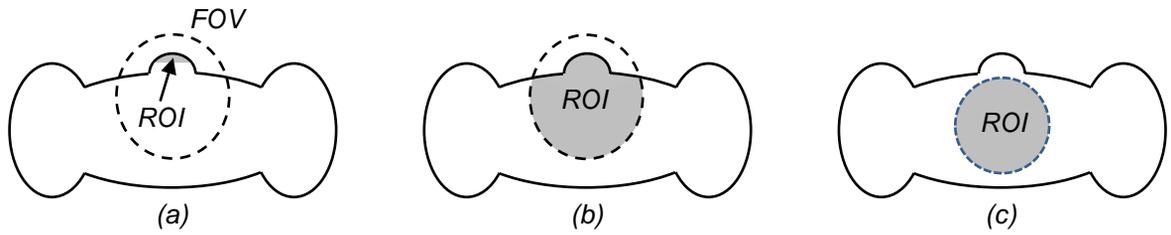

Figure 2. Exact ROI reconstruction conditions in comparison. For a field of view (FOV, a dashed ellipse) through which all line integrals are measured, exactly and stably recoverable ROIs (in grey) allowed by different data completeness conditions are not the same. (a) A small ROI permitted by Noo et al. [80], (b) an enlarged ROI enabled by Defrise et al. [85], (c) with interior tomography an ROI is as large as an FOV [16].

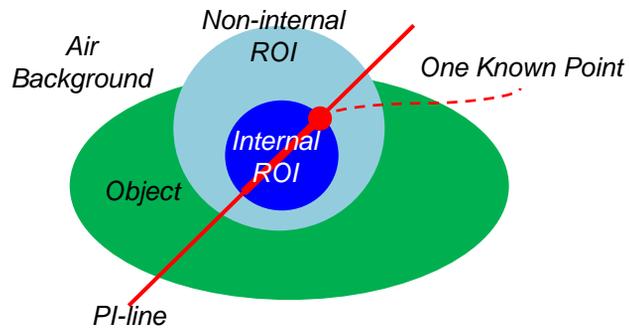

Figure 3. Conjecture for interior tomography – Can an internal ROI be exactly and stably reconstructed only from local data through the ROI assuming a known point? This conjecture led to known-subregion-based and sparsity-model-based interior tomography respectively.

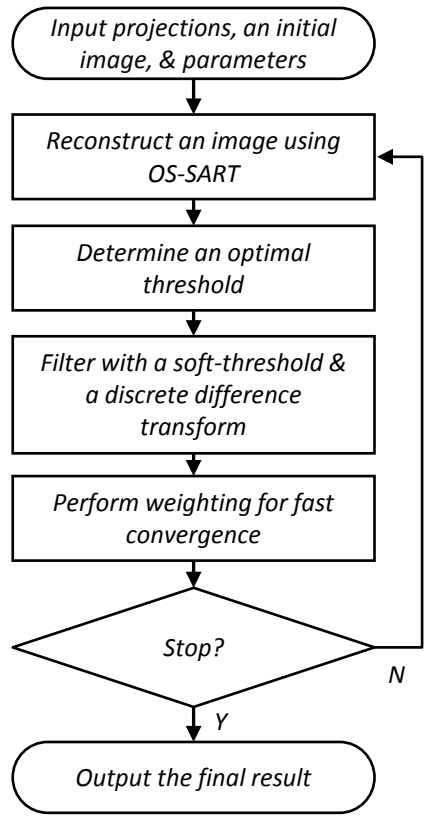

*Figure 4. Algorithmic flowchart for soft-threshold filtering based interior tomography.*

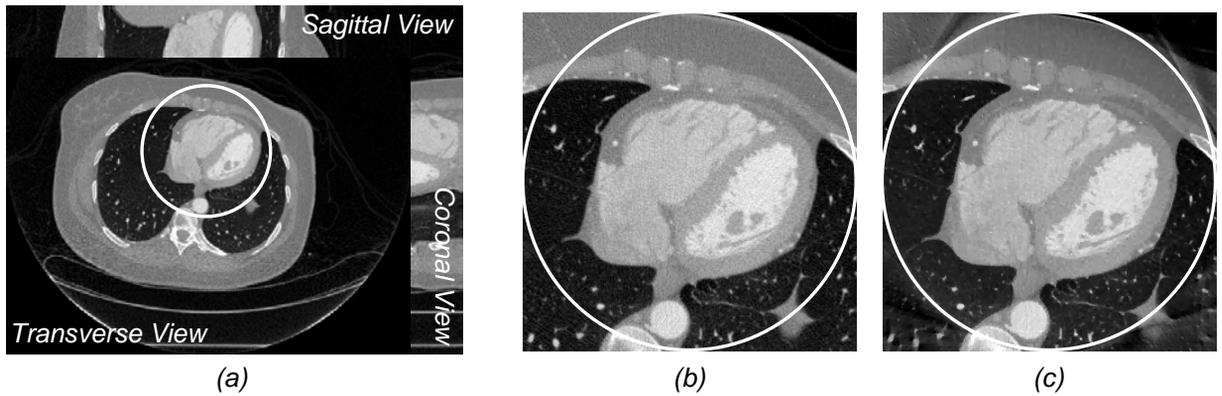

*Figure 5. Interior CT demonstration from a clinical CT scan. (a) A volumetric image reconstructed using the filtered backprojection method, (b) a magnified interior cardiac ROI in a transverse slice in (a), and (c) the interior ROI reconstruction corresponding to (b) using sparsity-model-based interior tomography. This figure is reprint from [88].*

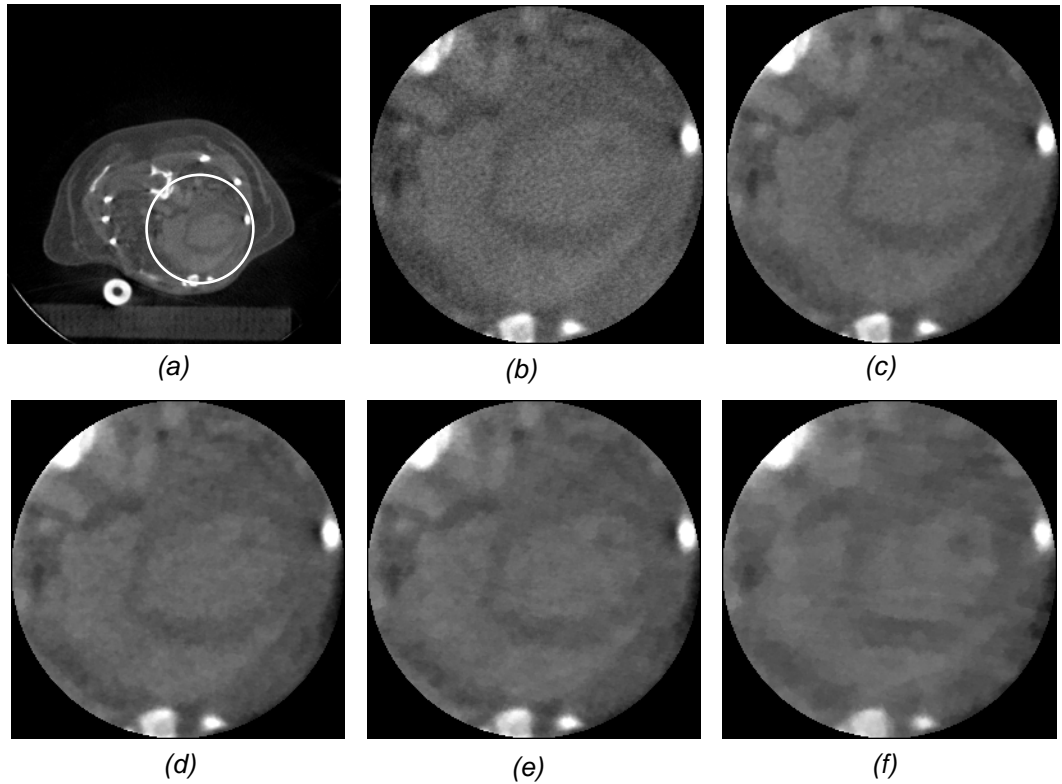

*Figure 6. Less is lower – Interior tomography reduces radiation dose [85]. (a) The image reconstructed from a global scan using a CNT-based micro-CT system at University of North Carolina, with the white circle for a cardiac ROI, (b) the local magnification of the ROI in (a), and (c) a sparsity-model-based interior micro-CT reconstruction from 400 projections after 60 iterations (without precise knowledge of any subregion in the ROI), (d)-(f) images reconstructed from 200, 100 and 50 projections to reduce radiation dose to 50%, 25% and 12.5% of that for (c), respectively. This figure is reprint from [95].*

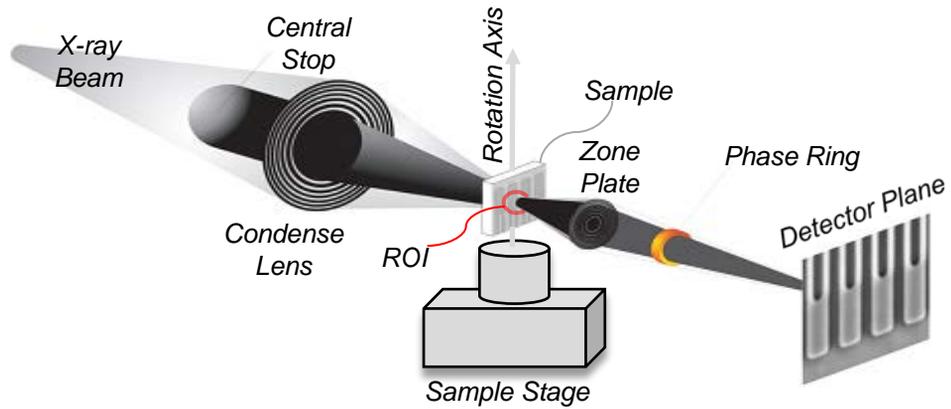

*Figure 7. Less is larger – Interior nano-CT promises to handle a sample larger than the x-ray beam. Current sample reduction into a fixed x-ray beam is tedious and error-prone. This is a collaborative project between Biomedical Imaging Division at Virginia Tech and Xradia (http://www.xradia.com).*

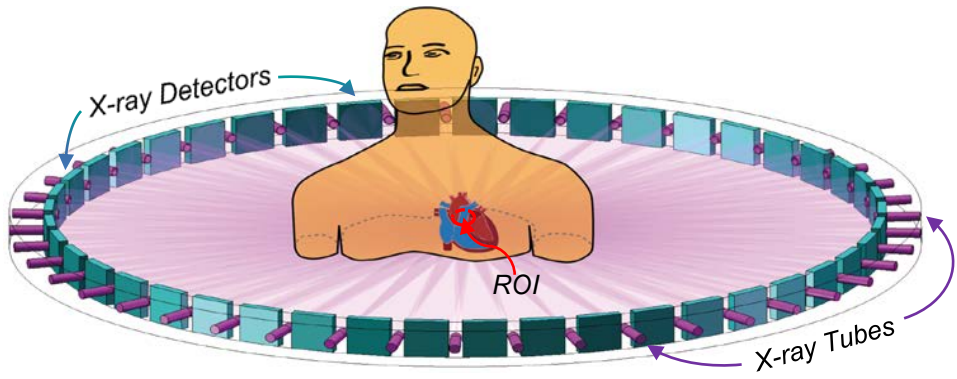

*Figure 8. Less is faster – Interior tomography accelerates data acquisition. Since interior tomography allows uses a small detector, a multi-source interior tomography scheme can be used for parallel data acquisition. In the limiting case, instantaneous tomographic imaging of a small ROI becomes feasible [20].*

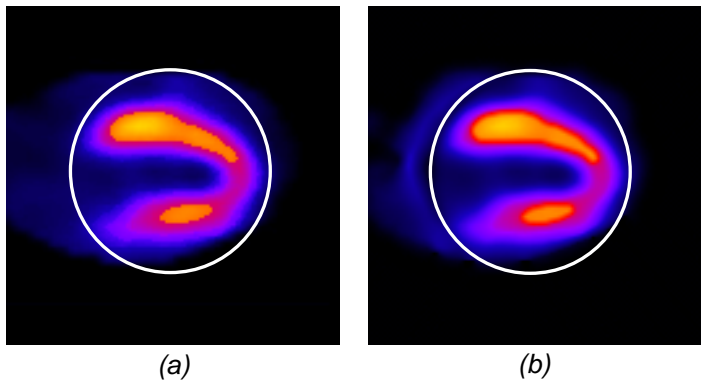

*Figure 9. Interior SPECT demonstration with a numerical cardiac phantom image. (a) An SPECT cardiac ROI image of 128 by 128 pixels, and (b) an interior ROI reconstruction assuming the attenuation coefficient $\mu_0$=0.15 after 40 iterations .*

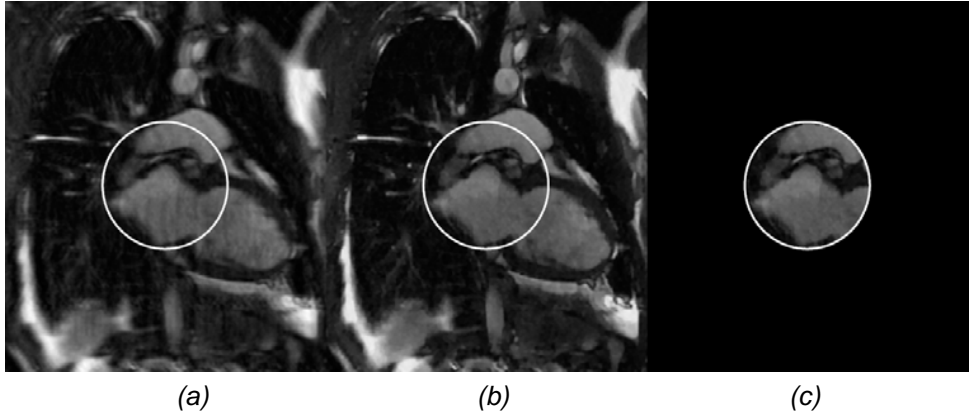

*Figure 10. Interior MRI demonstration with a numerical cardiac phantom image. (a) A global image reconstructed using the inverse fast Fourier transform (IFFT) method from randomly under-sampled MRI data (25%) along the phase-encoding direction, (b) the counterpart of (a) using the total variation (TV) minimization method, and (c) an interior MRI reconstruction using the TV minimization method from interior MRI data. This figure is reprint from [112].*

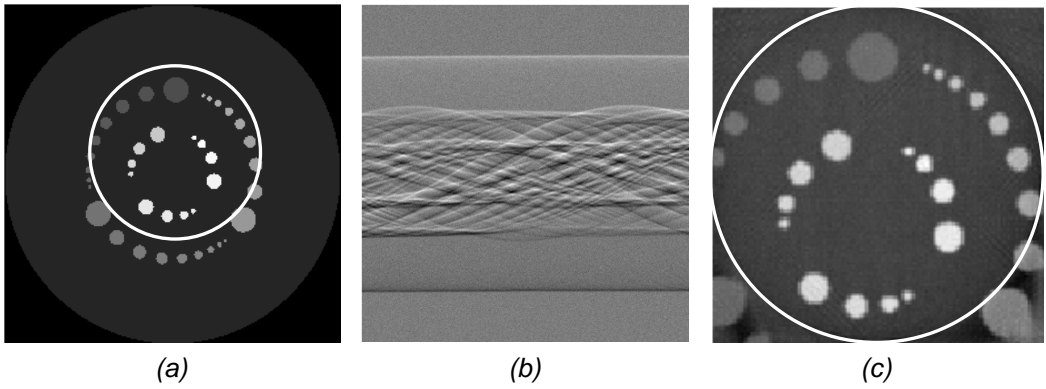

*Figure 11. Interior phase-contrast tomography demonstration with a numerical phantom. (a) The original numerical phantom, (b) a truncated differential phase-contrast sinogram simulated assuming a grating-based x-ray interferometer setup, and (c) the reconstructed image using interior differential phase-contrast tomography.*

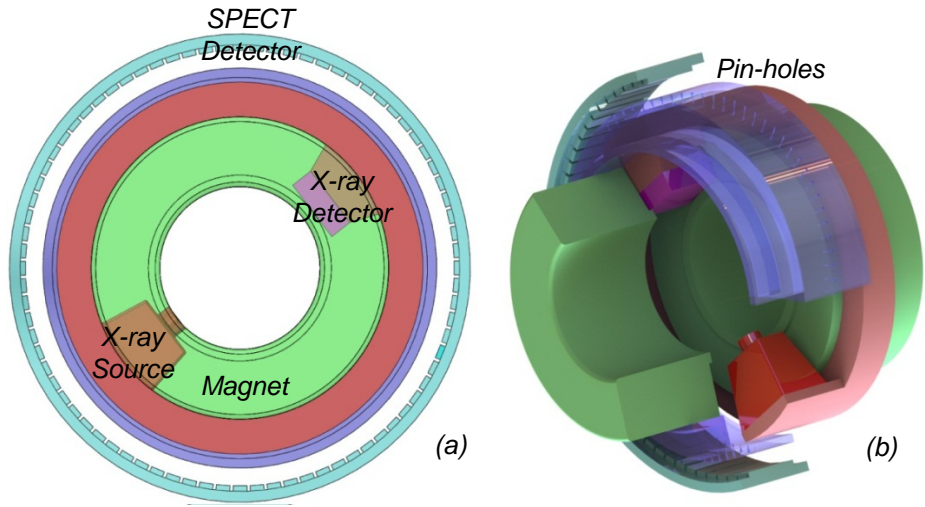

*Figure 12. Omni-tomography idea for grand fusion of multiple imaging modalities. (a) and (b) are cross-sectional and volumetric views respectively, where two donut-shaped magnets are paired to define a uniform background field covering an intended ROI, while other modalities such as CT and SPECT can be arranged between the magnetic rings [99]. To facilitate electromagnetic shielding, CT and and/or SPECT subsystems can be made stationary with a multi-source and/or multi-camera interior imaging arrangement (patent pending).*

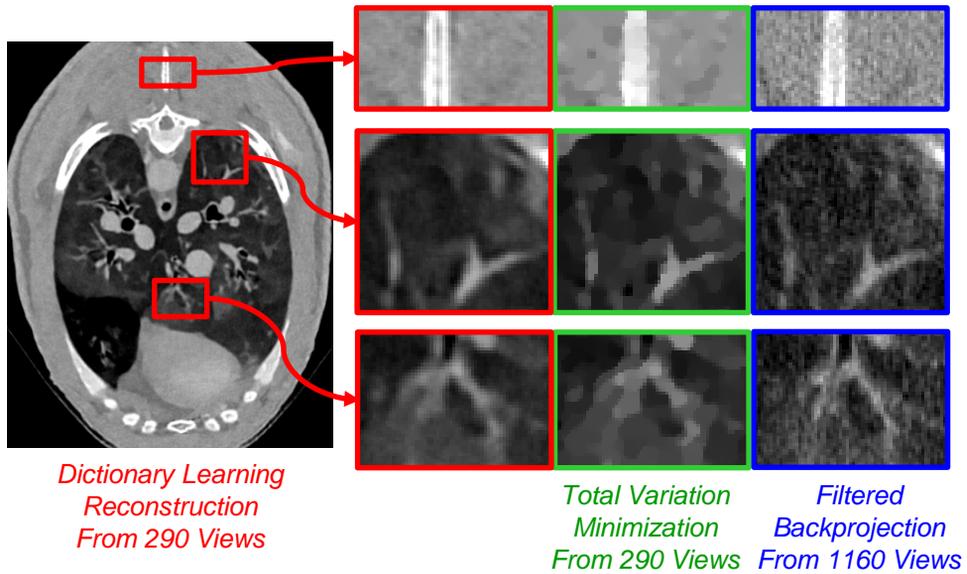

*Figure 13. Dictionary-learning-based reconstruction outperforming TV-minimization-based reconstruction [124]. The more knowledge we have, the less amount of data we need for satisfactory reconstruction. Dictionary learning promises to extract prior knowledge flexibly and effectively from training images, being more specific than a TV-based sparsifying constraint or a generic wavelet transform.*

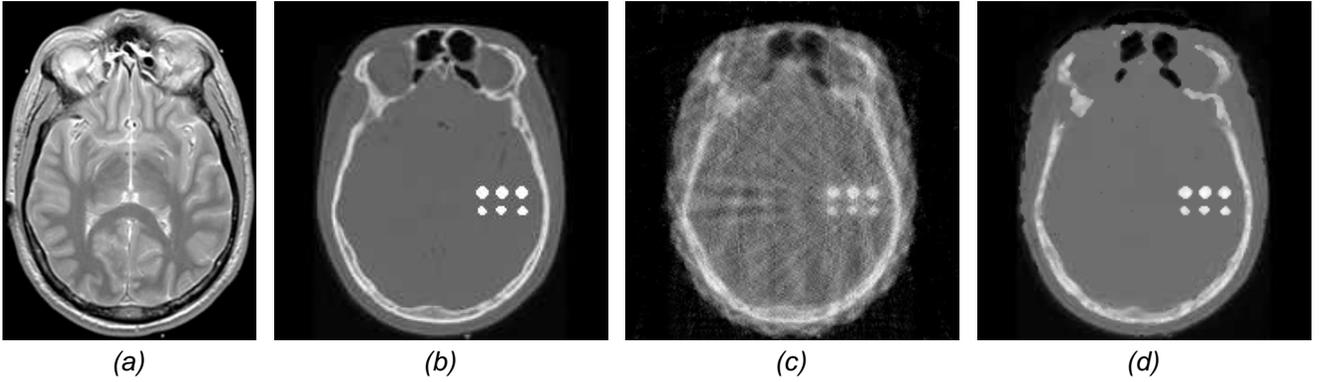

*Figure 14. Idea for omni-tomographic reconstruction in a unified framework. (a) an MR image phantom, (b) a CT image phantom corresponding to (a), where six implanted calcified regions are embedded, (c) a traditional CT reconstruction using the filtered backprojection method from 17 projections, and (d) dual-dictionary-learning-based CT reconstruction from the same 17 projections but aided by MRI data synthesized from (a) with 4-fold under-sampling. This figure is reprint from [139].*